\documentclass[11pt]{article}

\usepackage{a4wide}
\usepackage{times}
\usepackage{natbib}
\usepackage{amssymb}
\usepackage{amsmath}
\usepackage{graphicx}
\usepackage{lineno}
\usepackage{amsfonts}
\usepackage{rotating}

\def\var{\text{Var}}
\def\cov{\text{Cov}}

\begin{document}

%\begin{frontmatter}

\title{Flexible least squares for temporal data mining \\ and statistical arbitrage}

%\author{
%\begin{tabular}[t] {c@{\extracolsep{2em}}cc}
%Giovanni Montana$^\dag$ & Kostas Triantafyllopoulos$^\ddag$ & Theodoros Tsagaris$^\dag$\footnote{The author is also
%affiliated with BlueCrest Capital Management. The views presented here reflect solely the author's opinion.} \\
%\end{tabular}\\
%\begin{tabular}[t] {c@{\extracolsep{2em}}c}
%\it        $^\dag$Department of Mathematics & \it $^\ddag$Department of Probability and Statistics \\
%\it        Imperial College & \it University of Sheffield \\
%\it        London SW7 2AZ, UK & \it Sheffield S3 7RH, UK \\
%\end{tabular}
%}

\author{Giovanni Montana\footnote{Imperial College London, Department of Mathematics}, Kostas Triantafyllopoulos \footnote{University of Sheffield, Department of Probability and Statistics} , Theodoros Tsagaris \footnote{Imperial College London, Department of Mathematics, and BlueCrest Capital Management. The views presented here reflect solely the author's opinion.}  }

%\author{Giovanni Montana }
%\ead{g.montana@imperial.ac.uk} \corauth[cor]{Corresponding author: }

%\author{Kostas Triantafyllopoulos}

%\author{Theodoros Tsagaris \footnote{The author is also affiliated with BlueCrest Capital Management. The views presented here reflect solely the author's
%opinion.}}

%\address[IC]{Department of Mathematics, Statistics Section \\ Imperial College London \\ London SW7 2AZ, UK}

%\address[SU]{Department of Probability and Statistics \\ University of Sheffield \\  Sheffield S3 7RH, UK}

\maketitle

\begin{abstract}

A number of recent emerging applications call for studying data streams, potentially infinite flows of information
updated in real-time. When multiple co-evolving data streams are observed, an important task is to determine how these
streams depend on each other, accounting for dynamic dependence patterns without imposing any restrictive
probabilistic law governing this dependence. In this paper we argue that flexible least squares (FLS), a penalized
version of ordinary least squares that accommodates for time-varying regression coefficients, can be deployed
successfully in this context. Our motivating application is statistical arbitrage, an investment strategy that
exploits patterns detected in financial data streams. We demonstrate that FLS is algebraically equivalent to the
well-known Kalman filter equations, and take advantage of this equivalence to gain a better understanding of FLS and
suggest a more efficient algorithm. Promising experimental results obtained from a FLS-based algorithmic trading
system for the S\&P $500$ Futures Index are reported.

\end{abstract}

%\begin{keywords}
%Temporal data mining \sep flexible least squares \sep time-varying regression \sep algorithmic trading system \sep
%statistical arbitrage
%\end{keyword}

\emph{Keywords}: Temporal data mining, flexible least squares, time-varying regression, algorithmic trading system,
statistical arbitrage

%\end{frontmatter}

% \appendix

% \section{}
% \label{}

% Bibliographic references with the natbib package:
% Parenthetical: \citep{Bai92} produces (Bailyn 1992).
% Textual: \citet{Bai95} produces Bailyn et al. (1995).
% An affix and part of a reference:
%   \citep[e.g.][Ch. 2]{Bar76}
%   produces (e.g. Barnes et al. 1976, Ch. 2).

\section{Introduction} \label{intro}

Temporal data mining is a fast-developing area concerned with
processing and analyzing high-volume, high-speed data streams. A
common example of data stream is a time series, a collection of
univariate or multivariate measurements indexed by time.
Furthermore, each record in a data stream may have a complex
structure involving both continuous and discrete measurements
collected in sequential order. There are several application areas
in which temporal data mining tools are being increasingly used,
including finance, sensor networking, security, disaster management,
e-commerce and many others. In the financial arena, data streams are
being monitored and explored for many different purposes such as
algorithmic trading, smart order routing, real-time compliance, and
fraud detection. At the core of all such applications lies the
common need to make time-aware, instant, intelligent decisions that
exploit, in one way or another, patterns detected in the data.

In the last decade we have seen an increasing trend by investment
banks, hedge funds, and proprietary trading boutiques to systematize
the trading of a variety of financial instruments. These companies
resort to sophisticated trading platforms based on predictive models
to transact market orders that serve specific speculative investment
strategies.

Algorithmic trading, otherwise known as automated or systematic
trading, refers to the use of expert systems that enter trading
orders without any user intervention; these systems decide on all
aspects of the order such as the timing, price, and its final
quantity. They effectively implement pattern recognition methods in
order to detect and exploit market inefficiencies for speculative
purposes. Moreover, automated trading systems can slice a large
trade automatically into several smaller trades in order to hide its
impact on the market (a technique called \emph{iceberging}) and
lower trading costs. According to the Financial Times, the London
Stock Exchange foresees that about $60\%$ of all its orders in the
year 2007 will be entered by algorithmic trading.

Over the years, a plethora of statistical and econometric techniques
have been developed to analyze financial data \citep{DeGooijer2006}.
Classical time series analysis models, such as ARIMA and GARCH, as
well as many other extensions and variations, are often used to
obtain insights into the mechanisms that generates the observed data
and make predictions \citep{Chatfield04}. However, in some cases,
conventional time series and other predictive models may not be up
to the challenges that we face when developing modern algorithmic
trading systems. Firstly, as the result of developments in data
collection and storage technologies, these applications generate
massive amounts of data streams, thus requiring more efficient
computational solutions. Such streams are delivered in real time; as
new data points become available at very high frequency, the trading
system needs to quickly adjust to the new information and take
almost instantaneous buying and selling decisions. Secondly, these
applications are mostly exploratory in nature: they are intended to
detect patterns in the data that may be continuously changing and
evolving over time. Under this scenario, little prior knowledge
should be injected into the models; the algorithms should require
minimal assumptions about the data-generating process, as well as
minimal user specification and intervention.

In this work we focus on the problem of identifying time-varying
dependencies between co-evolving data streams. This task can be
casted into a regression problem: at any specified point in time,
the system needs to quantify to what extent a particular stream
depends on a possibly large number of other explanatory streams. In
algorithmic trading applications, a data stream may comprise daily
or intra-day prices or returns of a stock, an index or any other
financial instrument. At each time point, we assume that a target
stream of interest depends linearly on a number of other streams,
but the coefficients of the regression models are allowed to evolve
and change smoothly over time.

The paper is organized as follows. In section \ref{statarb} we
briefly review a number of common trading strategies and formulate
the problem arising in \emph{statistical arbitrage}, thus proving
some background material and motivation for the proposed methods.
The flexible least squares (FLS) methodology is introduced in
Section \ref{fls} as a powerful exploratory method for temporal data
mining; this method fits our purposes well because it imposes no
probabilistic assumptions and relies on minimal parameter
specification. In Section \ref{KF} some assumptions of the FLS
method are revisited, and we establish a clear connection between
FLS and the well-known Kalman filter equations. This connection
sheds light on the interpretation of the model, and naturally yields
a modification of the original FLS that is computationally more
efficient and numerically stable. Experimental results that have
been obtained using the FLS-based trading system are described in
Section \ref{sec:system}. In that section, in order to deal with the
large number of predictors, we complement FLS with a feature
extraction procedure that performs on-line dimensionality reduction.
We conclude in Section \ref{conclusions} with a discussion on
related work and directions for further research.

\section{A concise review of trading strategies} \label{statarb}

Two popular trading strategies are \emph{market timing} and
\emph{trend following}. Market timers and trend followers both
attempt to profit from price movements, but they do it in different
ways. A market timer forecasts the direction of an asset, going long
(i.e. buying) to capture a price increase, and going short (i.e.
selling) to capture a price decrease. A trend follower attempts to
capture the market trends. Trends are commonly related to serial
correlations in price changes; a trend is a series of asset prices
that move persistently in one direction over a given time interval,
where price changes exhibit positive serial correlation. A trend
follower attempts to identify developing price patterns with this
property and trade in the direction of the trend if and when this
occurs.

Although the time-varying regression models discussed in this work
may be used to implement such trading strategies, we will not
discuss this further. We rather focus on \emph{statistical
arbitrage}, a class of strategies widely used by hedge funds or
proprietary traders. The distinctive feature of such strategies is
that profits can be made by exploiting statistical \emph{mispricing}
of one or more assets, based on the expected value of these assets.

The simplest special case of these strategies is perhaps \emph{pairs
trading} (see \cite{Elliott2005,Gatev2006}). In this case, two
assets are initially chosen by the trader, usually based on an
analysis of historical data or other financial considerations. If
the two stocks appear to be tied together in the long term by some
common stochastic trend, a trader can take maximum advantage from
temporary deviations from this assumed equilibrium \footnote{This
strategy relies on the idea of \emph{co-integration}. Several
applications of cointegration-based trading strategies are presented
in \cite{Alexander2002} and \cite{Burgess2003}.}.

\begin{figure}
\centering
\includegraphics[width=10cm, height=7cm]{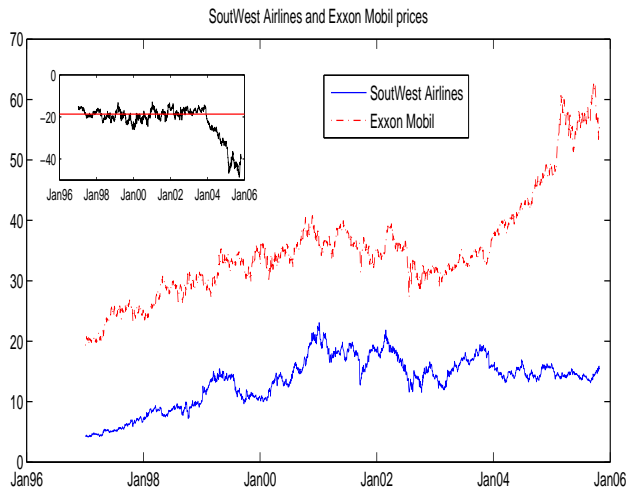}
\caption{{\small Historical prices of Exxon Mobil Corporation and
SouthWest Airlines for the period 1997-2007. The spread time series,
reported in the inset, shows an equilibrium level between the two
prices until about January 2004.}} \label{fig:pairsexample}
\end{figure}

A specific example will clarify this simple but effective strategy.
Figure \ref{fig:pairsexample} shows the historical prices of two
assets, SouthWest Airlines and Exxon Mobil; we denote the two price
time series by $y_t$ and $x_t$ for $t=1,2,\dots$, respectively.
Clearly, from $1997$ till $2004$, the two assets exhibited some
dependence: their spread, defined as $s_t=y_t-x_t$ (plotted in the
inset figure) fluctuates around a long-term average of about $-20$.
A trading system implementing a pairs trading strategy on these two
assets would exploit temporary divergences from this market
equilibrium. For instance, when the spread $s_t$ is greater than
some predetermined positive constant $c$, the system assume that the
SouthWest Airlines is overpriced and would go short on SouthWest
Airlines and long on Exxon Mobil, in some predetermined ratio. A
profit is made when the prices revert back to their long-term
average. Although a stable relationship between two assets may
persist for quite some time, it may suddenly disappear or present
itself in different patterns, such as periodic or trend patterns. In
Figure \ref{fig:pairsexample}, for instance, the spread shows a
downward trend after January $2004$, which may be captured by
implementing more refined models.

\subsection{A statistical arbitrage strategy}

Opportunities for pairs trading in the simple form described above
are dependent upon the existence of similar pairs of assets, and
thus are naturally limited. Many other variations and extensions
exist that exploit temporary mispricing among securities. For
instance, in \emph{index arbitrage}, the investor looks for
temporary discrepancies between the prices of the stocks comprising
an index and the price of a futures contract\footnote{A futures
contract is an obligation to buy or sell a certain underlying
instrument at a specific date and price, in the future.}on that
index. By buying either the stocks or the futures contract and
selling the other, market inefficiency can be exploited for a
profit.

In this paper we adopt a simpler strategy than index arbitrage,
somewhat more related to pairs trading. The trading system we
develop tries to exploit discrepancies between a \emph{target
asset}, selected by the investor, and a paired \emph{artificial
asset} that reproduces the target asset. This artificial asset is
represented by a data stream obtained as a linear combination of a
possibly large set of \emph{explanatory} streams assumed to be
correlated with the target stream.

The rationale behind this approach is the following: if there is a
strong association between synthetic and target assets persisting
over a long period of time, this association implies that both
assets react to some underlying (and unobserved) systematic
component of risk that explains their dynamics. Such a systematic
component may include all market-related sources of risk, including
financial and economic factors. The objective of this approach is to
neutralize all marker-related sources of risks and ultimately obtain
a data stream that best represents the target-specific risk, also
known as \emph{idiosyncratic} risk.

Suppose that $y_t$ represents the data stream of the target asset,
and $\widehat y_t$ is the artificial asset estimated using a set of
$p$ explanatory and co-evolving data streams $x_1, \ldots, x_p$,
over the same time period. In this context, the artificial asset can
also be interpreted as the \emph{fair price} of the target asset,
given all available information and market conditions. The
difference $y_t-\widehat y_t$ then represents the risk associated
with the target asset only, or \emph{mispricing}. Given that this
construction indirectly accounts for all sources of variations due
to various market-related factors, the mispricing data stream is
more likely to contain predictable patterns (such as the
mean-reverting behavior seen in Figure \ref{fig:pairsexample}) that
could potentially be exploited for speculative purposes. For
instance, in an analogy with the pairs trading approach, a possibly
large mispricing (in absolute value) would flag a temporary
inefficiency that will soon be corrected by the market. This
construction crucially relies on accurately and dynamically
estimating the artificial asset, and we discuss this problem next.

\section{Flexible Least Squares (FLS) } \label{fls}

The standard linear regression model involves a response variable
$y_t$ and $p$ predictor variables $x_{1},\ldots,x_{p}$, which
usually form a predictor column vector
$x_t=(x_{1t},\ldots,x_{pt})'$. The model postulates that $y_t$ can
be approximated well by $x_t' \beta$, where $\beta$ is a
$p$-dimensional vector of regression parameters. In ordinary least
square (OLS) regression, estimates $\widehat \beta$ of the parameter
vector are found as those values that minimize the cost
function
\begin{equation} \label{traditional_cost} C(\beta) =
\sum_{t=1}^T (y_t-x_t'\beta)^2
\end{equation}
%\begin{equation} \label{optimization}
%\widehat \beta = \min_{\beta} C(\beta)
%\end{equation}

When both the response variable $y_t$ and the predictor vector $x_t$
are observations at time $t$ of co-evolving data streams, it may be
possible that the linear dependence between $y_t$ and $x_t$ changes
and evolves, dynamically, over time. Flexible least squares were
introduced at the end of the 80's by \cite{Tesfatsion1989} as a
generalization of the standard linear regression model above in
order to allow for time-variant regression coefficients. Together
with the usual regression assumption that
\begin{equation} \label{eq:regression} y_t-x_t' \beta_t \approx
0
\end{equation}
the FLS model also postulates that
\begin{equation} \label{assumption}
\beta_{t+1}-\beta_t \approx 0
\end{equation}
that is, the regression coefficients are now allowed to evolve
slowly over time.

FLS does not require the specification of probabilistic properties
for the residual error in \eqref{eq:regression}. This is a favorable
aspect of the method for applications in temporal data mining, where
we are usually unable to precisely specify a model for the errors;
moreover, any assumed model would not hold true at all times. We
have found that FLS performs well even when assumption
\eqref{assumption} is violated, and there are large and sudden
changes from $\beta_{t-1}$ to $\beta_t$, for some $t$. We will
illustrate this point by means of an example in the next section.

With these minimal assumptions in place, given a predictor $x_t$, a
procedure is called for the estimation of a unique path of
coefficients, $\beta_t=(\beta_{1t}',\ldots,\beta_{pt}')'$, for $t=1,
2, \ldots$. The FLS approach consists of minimizing a penalized
version of the OLS cost function \eqref{traditional_cost},
namely\footnote{This cost function is called the
\emph{incompatibility cost} in \cite{Tesfatsion1989}}
\begin{equation} \label{original_cost}
C(\beta; \mu) = \sum_{t=1}^T (y_t-x_t'\beta_t)^2 + \mu
\sum_{t=1}^{T-1} \xi_t
\end{equation}
where we have defined
\begin{equation} \label{eq:bt}
\xi_t = (\beta_{t+1}-\beta_t)'(\beta_{t+1}-\beta_t)
\end{equation}
and $\mu \geq 0$ is a scalar to be determined.

In their original formulation, \cite{Kalaba1988} propose an
algorithm that minimizes this cost with respect to every $\beta_t$
in a sequential way. They envisage a situation where \emph{all} data
points are stored in memory and promptly accessible, in an off-line
fashion. The core of their approach is summarized in the sequel for
completeness.

The smallest cost of the estimation process at time $t$ can be
written recursively as
\begin{equation} \label{eq:smallestcost}
c(\beta_{t+1};\mu) = \inf_{\beta_t} \left\{ (y_t-x_t'\beta_t)^2+ \mu \xi_t + c(\beta_t; \mu) \right\}
\end{equation}
Furthermore, this cost is assumed to have a quadratic form
\begin{equation} \label{eq:quadratic}
c(\beta_t; \mu)=\beta_t' S_{t-1}\beta_t - 2\beta_t' s_{t-1} +r_{t-1}
\end{equation}
where $S_{t-1}$ and $s_{t-1}$ have dimensions $p \times p$ and $p
\times 1$, respectively, and $r_{t-1}$ is a scalar. Substituting
\eqref{eq:quadratic} into \eqref{eq:smallestcost} and then
differentiating the cost \eqref{eq:smallestcost} with respect to
$\beta_t$, conditioning on $\beta_{t+1}$, one obtains a recursive
updating equation for the time-varying regression coefficient
\begin{equation} \label{eq:recursive}
\widehat \beta_t = d_t + M_t \beta_{t+1}
\end{equation}
with
\begin{align*}
d_t  & = \mu^{-1} M_t (s_{t-1} + x_t y_t) \\
M_t & = \mu (S_{t-1} + \mu I_p + x_t x_t')^{-1}
\end{align*}
The recursions are started with some initial $S_0$ and $s_0$. Now,
using \eqref{eq:recursive}, the cost function can be written as
$$
c(\beta_{t+1}; \mu) = \beta_{t+1}' S_{t+1} - 2\beta_{t+1}' s_t+r_t
$$
where
\begin{align}
S_t & = \mu(I_p-M_t) \label{St:1} \\
s_t & = \mu d_t \label{St:2} \\
r_t & = r_{t-1} + y_t^2 - (s_{t-1}+x_t y_t)'d_t \nonumber
\end{align}
and where $I_p$ is the $p\times p$ identity matrix. In order to
apply \eqref{eq:recursive}, this procedure requires all data points
till time $T$ to be available, so the coefficient vector $\beta_T$
should be computed first. \cite{Kalaba1988} show that the estimate
of $\beta_T$ can be obtained sequentially as
$$
\widehat \beta_T = (S_{T-1} + x_T x_T')^{-1}(s_{T-1}+x_T y_T)
$$
Subsequently, \eqref{eq:recursive} can be used to estimate all
remaining coefficient vectors $\beta_{T-1},\ldots,\beta_1$, going
backwards in time.

The procedure relies on the specification of the regularization
parameter $\mu \geq 0$; this scalar penalizes the dynamic component
of the cost function \eqref{original_cost}, defined in
\eqref{eq:bt}, and acts as a smoothness parameter that forces the
time-varying vector towards or away from the fixed-coefficient OLS
solution. We prefer the alternative parameterization based on
$\mu=(1-\delta)/\delta$ controlled by a scalar $\delta$ varying in
the unit interval. Then, with $\delta$ set very close to $0$
(corresponding to very large values of $\mu$), near total weight is
given to minimizing the static part of the cost function
\eqref{original_cost}. This is the smoothest solution and results in
standard OLS estimates. As $\delta$ moves away from $0$, greater
priority is given to the dynamic component of the cost, which
results in time-varying estimates.

\subsection{Off-line and on-line FLS: an illustration}

As noted above, the original FLS has been introduced for situations
in which all the data points are available, in batch, prior to the
analysis. In contrast, we are interested in situations where each
data point arrives sequentially. Each component of the $p$
dimensional vector $x_t$ represents a new point of a data stream,
and the path of regression coefficients needs to be updated at each
time step so as to incorporate the most recently acquired
information. Using the FLS machinery in this setting, the estimate
of $\beta_t$ is given recursively by
\begin{equation}
\widehat{\beta}_t = (S_{t-1}+x_tx_t')^{-1}(s_{t-1}+x_ty_t)
\label{eq:Betat_1}
\end{equation}
where, by substituting $M_t$ and $d_t$ in \eqref{St:1} and
\eqref{St:2}, we obtain the recursions of $S_t$ and $s_t$ as
\begin{gather}
S_t=\mu (S_{t-1}+\mu I_p +x_tx_t')^{-1} (S_{t-1}+x_tx_t')\label{eq:St} \\
s_t = \mu (S_{t-1}+\mu I_p +x_tx_t')^{-1}(s_{t-1}+x_ty_t)\nonumber
\end{gather}
These recursions are initially started with some arbitrarily chosen
values $S_0$ and $s_0$.

\begin{figure}
\centering
\includegraphics[width=10cm, height=7cm]{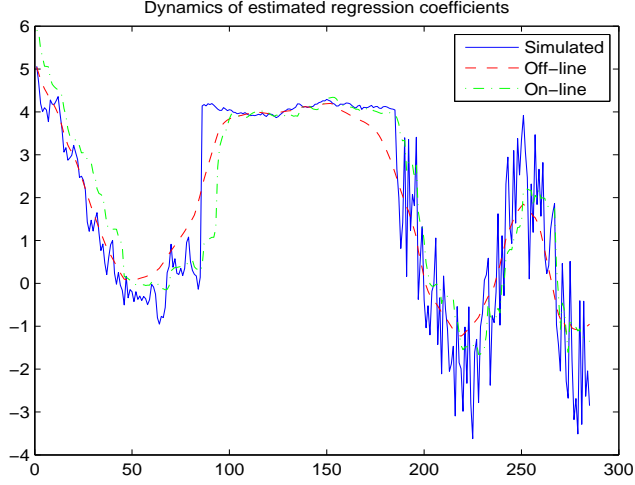}
\caption{{\small Simulated versus estimated time-varying regression coefficients
using FLS in both off-line and on-line mode.}} \label{fig:flsexample}
\end{figure}

Figure \ref{fig:flsexample} illustrates how accurately the FLS algorithm
recovers the path of the time-varying coefficients, in both off-line
and on-line settings, for some artificially created data streams. The target
stream $y_t$ for this example has been generated using the model
\begin{equation} \label{eq:simulation}
y_t = x_t \beta_t + \epsilon_t
\end{equation}
where $\epsilon_t$ is uniformly distributed over the interval $[-2,2]$ and
the explanatory stream $x_t$ evolves as
$$
x_t = 0.8 x_{t-1} + z_t
$$
with $z_t$ being white noise. The regression coefficients have been
generated using a slightly complex mechanism for the purpose of
illustrating the flexibility of FLS. Starting with $\beta_1=7$, we
then generate $\beta_t$ as
\begin{equation*}
\beta_t=
\begin{cases}
\beta_{t-1}+a_t & \quad \text{ for } t=2,\ldots,99 \\
\beta_{t-1}+4 & \quad \text{ for } t=100 \\
\beta_{t-1}+b_t & \quad \text{ for } t=101,\ldots,200 \\
5\sin(0.5t)+c_t & \quad \text{ for } t=201,\ldots,300 \\
\end{cases}
\end{equation*}
where $a_t$ and $b_t$ are Gaussian random variables with standard
deviations $0.1$ and $0.001$, respectively, and $c_t$ is uniformly
distributed over $[-2,2]$. We remark that this example features
non-Gaussian error terms, as well as linear and non-linear behaviors
in the dynamics of the regression coefficient, varying over time.

In this example we set $\delta=0.98$. Although such a high value of
$\delta$ encourages the regression parameters to be very dynamic,
the nearly constant coefficients observed between $t=101$ and
$t=200$, as well as the two sudden jumps at times $t=100$ and
$t=201$, are estimated well, and especially so in the on-line
setting. The non-linear dynamics observed from time $t=201$ onwards
is also well captured.

\section{An alternative look at FLS} \label{KF}

In section \ref{fls}, we have stressed that FLS relies on a quite
general assumption concerning the evolution of the regression
coefficients, as it only requires $\beta_{t+1}-\beta_t$ to be small
at all times. Accordingly, assumption \eqref{assumption} does not
imply or require that each vector $\beta_t$ is a random vector.
Indeed, in the original work of \cite{Kalaba1988}, $\{\beta_t\}$ is
not treated as a sequence of random variables, but rather taken as a
sequence of unknown quantities to be estimated.

We ask ourselves whether we can gain a better understanding of the
FLS method after assuming that the regression coefficients are
indeed random vectors, without losing the generality and flexibility
of the original FLS method. As it turns out, if we are willing to
make such an assumption, it is possible to establish a neat
algebraic correspondence between the FLS estimation equations and
the well-known Kalman filter (KF) equations. This correspondence has
a number of advantages. Firstly, this connection sheds light into
the meaning and interpretation of the smoothing parameter $\mu$ in
the cost function \eqref{original_cost}. Secondly, once the
connection with KF is established, we are able to estimate the
covariance matrix of the estimator of $\beta_t$. Furthermore, we are
able to devise a more efficient version of FLS that does not require
any matrix inversion. As in the original method, we restrain from
imposing any specific probability distribution. The reminder of this
section is dedicated to providing an alternative perspective of FLS,
and deriving a clear connection between this method and the
well-known Kalman filter equations.

\subsection{The state-space model}

In our formulation, the regression coefficient at time $t+1$ is
modeled as a noisy version of the previous coefficient at time $t$.
First, we introduce a random vector $\omega_t$ with zero mean and
some covariance matrix $V_\omega$, so that
\begin{equation}\label{eq:ss1}
\beta_{t+1}=\beta_t+\omega_t\quad t=0,1,\ldots,T-1.
\end{equation}
Then, along the same lines, we introduce a random variable
$\epsilon_t$ having zero mean and some variance $V_\epsilon$, so
that
\begin{equation}\label{eq:ss2}
y_t=x_t'\beta_t+\epsilon_t\quad t=1,\ldots,T.
\end{equation}
Equations (\ref{eq:ss1}) and (\ref{eq:ss2}), jointly considered,
result in a linear state-space model, for which it is assumed that
the innovation series $\{\epsilon_t\}$ and $\{\omega_t\}$ are
mutually and individually uncorrelated, i.e. $\epsilon_i$ is
uncorrelated of $\epsilon_j$, $\omega_i$ is uncorrelated of
$\omega_j$, and $\epsilon_k$ is uncorrelated of $\omega_{\ell}$, for
any $i\neq j$ and for any $k,\ell$. It is also assumed that for all
$t$, $\epsilon_t$ and $\omega_t$ are uncorrelated of the initial
state $\beta_0$. It should be emphasized again that no specific
distribution assumptions for $\epsilon_t$ and $\omega_t$ have been
made. We only assume that $\epsilon_t$ and $\omega_t$ attain some
distributions, which we do not know. We only need to specify the
first two moments of such distributions. In this sense, the only
difference between the system specified by
(\ref{eq:ss1})-(\ref{eq:ss2}) and FLS is the assumption of
randomness of $\beta_t$.
%(the correspondence
%of $V_\epsilon$ and $V_\omega$ with $\mu$ will be defined later).

\subsection{The Kalman filter}

The Kalman filter \citep{KRE60} is a powerful method for the
estimation of $\beta_t$ in the above linear state-space model. In
order to establish the connection between FLS and KF, we derive an
alternative and self-contained proof of the KF recursions that make
no assumptions on the distributions of $\epsilon_t$ and $\omega_t$.
We have found related proofs of such recursions that do not rely on
probabilistic assumptions, as in \cite{KRE60} and \cite{ERL06}. In
comparison with these, we believe that our derivation is simpler and
does not involve matrix inversions, which serves our purposes well.

We start with some definitions and notation. At time $t$, we denote
by $\widehat{\beta}_t$ the estimate of $\beta_t$ and by
$\widehat{y}_{t+1}=E(y_{t+1})$ the one-step forecast of $y_{t+1}$,
where $E(.)$ denotes expectation. The variance of $y_{t+1}$ is known
as the one-step forecast variance and is denoted by
$Q_t=\var(y_{t+1})$. The one-step forecast error is defined as
$e_t=y_t-E(y_t)$. We also define the covariance matrix of
$\beta_t-\widehat{\beta}_t$ as $P_t$ and the covariance matrix of
$\beta_t-\widehat{\beta}_{t-1}$ as $R_t$ and we write
$\cov(\beta_t-\widehat{\beta}_t)=P_t$ and
$\cov(\beta_t-\widehat{\beta}_{t-1})=R_t$. With these definitions,
and assuming linearity of the system, we can see that, at time $t-1$
\begin{align*}
R_t & = P_{t-1}+V_\omega \\
\widehat{y}_t & = x_t'\widehat{\beta}_{t-1} \\
Q_t &  = x_t'R_tx_t+V_\epsilon
\end{align*}
where $P_{t-1}$ and $\widehat{\beta}_{t-1}$ are assumed known. The KF gives
recursive updating equations for $P_t$ and $\widehat{\beta}_t$ as functions of
$P_{t-1}$ and $\widehat{\beta}_{t-1}$.

Suppose we wish to obtain an estimator of $\beta_t$ that is linear
in $y_t$, that is $\widehat{\beta}_t=a_t+K_ty_t$, for some $a_t$ and
$K_t$ (to be specified later). Then we can write
\begin{equation}\label{eq:ss:est1}
\widehat{\beta}_t=a_t^*+K_te_t
\end{equation}
with $e_t=y_t-x_t'\widehat{\beta}_{t-1}$. We will show that for some
$K_t$, if $\widehat{\beta}_t$ is required to minimize the sum of
squares
\begin{equation}\label{sumsq1}
C=\sum_{t=1}^T (y_t-x_t'\beta_t)^2
\end{equation}
then $a_t^*=\widehat{\beta}_{t-1}$. To prove this, write
$Y=(y_1,\ldots,y_T)'$, $X=(x_1',\ldots,x_T')'$,
$B=(\beta_1',\ldots,\beta_T')'$, $\mathcal{E}=(e_1,\ldots,e_T)'$ and
$$
K=\left(\begin{array}{cccc} K_1 & 0 & \cdots & 0 \\ 0 & K_2 & \cdots
& 0 \\ \vdots & \vdots & \ddots & \vdots \\ 0 & 0 & \cdots &
K_T\end{array}\right)
$$
Then we can write (\ref{sumsq1}) as
$$
C \equiv C(B)=(Y-XB)'(Y-XB)
$$ and
$\widehat{B}=A^*+K\mathcal{E}$, where
$A^*=((a_1^*)',\ldots,(a_T^*)')'$. We will show that $A^*=B^*$,
where $B^*=(\widehat{\beta}_0',\ldots,\widehat{\beta}_{T-1}')'$.
With the above $\widehat{B}$, the sum of squares can be written as
\begin{eqnarray*}
\mathcal{S}(\widehat{B})&=& (Y-XA^*-XK\mathcal{E})'(Y-XA^*-XK\mathcal{E}) \\
&=& (Y-XA^*)'(Y-XA^*) -2(Y-XA^*)'XK\mathcal{E} \\ && +
\mathcal{E}'K'X'XK\mathcal{E}
\end{eqnarray*}
which is minimized when $Y-XA^*$ is minimized or when $E(Y-XA^*)=0$,
leading to $A^*=B^*$ as required. Thus, $a_t^*=\widehat{\beta}_{t-1}$ and from (\ref{eq:ss:est1}) we
have
\begin{equation}\label{beta:est:ss}
\widehat{\beta}_t=\widehat{\beta}_{t-1}+K_te_t
\end{equation}
for some value of $K_t$ to be defined. From the definition of $P_t$, we have
that
\begin{eqnarray}
P_t &=&  \cov(\beta_t-
(\widehat{\beta}_{t-1}+K_t(x_t'\beta_t+\epsilon_t-x_t'\widehat{\beta}_{t-1})))
\nonumber
\\ &=& \cov(
(I_p-K_tx_t')(\beta_t-\widehat{\beta}_{t-1})-K_t\epsilon_t) \nonumber \\
&=& (I_p-K_tx_t')R_t(I_p-x_tK_t')+V_\epsilon K_t K_t' \nonumber \\
&=& R_t-K_tx_t'R_t-R_tx_tK_t' + Q_tK_tK_t'\label{eq:Pt1}
\end{eqnarray}

Now, we can choose $K_t$ that minimizes
$$
E(\beta_t-\widehat{\beta}_t)'(\beta_t-\widehat{\beta}_t)
$$
which is the same as minimizing the trace of $P_t$, and thus $K_t$ is the
solution of the matrix equation
$$
\frac{ \partial \textrm{trace}(P_t) } { \partial K_t } =
-2(x_t'R_t)'+2Q_tK_t=0
$$
where $\partial \textrm{trace}(P_t) / \partial K_t$ denotes the
partial derivative of the trace of $P_t$ with respect to $K_t$.
Solving the above equation we obtain $K_t=R_tx_t/Q_t$. The quantity
$K_t$, also known as the \emph{Kalman gain}, is optimal in the sense
that among all linear estimators $\widehat{\beta}_t$,
(\ref{beta:est:ss}) minimizes
$E(\beta_t-\widehat{\beta}_t)'(\beta_t-\widehat{\beta}_t)$. With
$K_t=R_tx_t/Q_t$, from (\ref{eq:Pt1}) the minimum covariance matrix
$P_t$ becomes
\begin{equation}\label{eq:Pt2} P_t=R_t-Q_tK_tK_t'
\end{equation}

The KF consists of equations (\ref{beta:est:ss}) and (\ref{eq:Pt2}),
together with
\begin{align*}
K_t & = R_tx_t/Q_t \\
R_t & =P_{t-1}+V_\omega \\
Q_t & =x_t'R_tx_t+V_\epsilon \qquad \text{and} \\
e_t & =y_t-x_t'\widehat{\beta}_{t-1}
\end{align*}
Initial values for $\widehat{\beta}_0$ and $P_0$ have to be placed; usually
we set $\widehat{\beta}_0=0$ and $P_0^{-1}=0$.

Note that from the recursions of $P_t$ and $R_t$ we have
\begin{equation}\label{Rt:1}
R_{t+1}=R_t-Q_tK_tK_t'+V_{\omega}
\end{equation}

\subsection{Correspondence between FLS and KF}

Traditionally, the KF equations are derived under the assumption
that $\epsilon_t$ and $\omega_t$ follow the normal distribution, as
in \cite{JAZ70}. This stronger distributional assumption allows the
derivation of the likelihood function. When the normal likelihood is
available, we note that its maximization is equivalent to minimizing
the quantity
$$
\sum_{t=1}^T (y_t-x_t'\beta_t)^2 + \frac{1}{V_\omega}
\sum_{t=1}^{T-1} \xi_t
$$
with respect to $\beta_1,\ldots,\beta_T$, where $\xi_t$ has been
defined in \eqref{eq:bt} (see \cite{JAZ70} for a proof). The above
expression is exactly the cost function \eqref{original_cost} with
$\mu$ replaced by $1/V_\omega$.

This correspondence can now be taken a step further: in a more
general setting, where no distributional assumptions are made, it is
possible to arrive to the same result. This is achieved by
rearranging equation \eqref{eq:Betat_1} in the form of
\eqref{beta:est:ss}, which is the KF estimator of $\beta_t$. First,
note that from (\ref{eq:St}) we can write
$$
(S_{t-1}+x_tx_t')^{-1}=\mu S_t^{-1} (S_{t-1}+\mu I_p+x_tx_t')^{-1}
$$ and
substituting to (\ref{eq:Betat_1}) we get
$\widehat{\beta}_t=S_t^{-1}s_t$. Thus we have
\begin{eqnarray*}
\widehat{\beta}_t-\widehat{\beta}_{t-1} &=&
S_t^{-1}s_t-S_{t-1}^{-1}s_{t-1} \\ &=&
(S_{t-1}+x_tx_t')^{-1}(s_{t-1}+x_ty_t)-S_{t-1}^{-1}s_{t-1} \\ &=&
S_{t-1}^{-1}x_ty_t - \frac{ S_{t-1}^{-1}x_tx_t'S_{t-1}^{-1}
(s_{t-1}+x_ty_t) } {x_t'S_{t-1}^{-1}x_t+1 } \\ &=& \frac{
S_{t-1}^{-1} x_t}{x_t'S_{t-1}^{-1}x_t+1} (y_tx_t'S_{t-1}^{-1}x_t+y_t \\
&& - x_t'S_{t-1}^{-1}s_{t-1}-x_t'S_{t-1}^{-1}x_ty_t) \\ &=& \frac{
S_{t-1}^{-1} x_t}{x_t'S_{t-1}^{-1}x_t+1}
(y_t-x_t'\widehat{\beta}_{t-1}) = K_te_t
\end{eqnarray*}
with
\begin{align*}
K_t & = R_tx_t/Q_t \\
R_t & = S_{t-1}^{-1} \\
Q_t & = x_t'R_tx_t+1 \\
V_\epsilon & =1
\end{align*}
It remains to prove that the recursion of $S_t$ as in (\ref{eq:St})
communicates with the recursion of \eqref{Rt:1}, for
$R_{t+1}=S_t^{-1}$. To end this, starting from (\ref{eq:St}) and
using the matrix inversion lemma, we obtain
\begin{eqnarray*}
R_{t+1}=S_t^{-1} &=& \mu^{-1} (S_{t-1}+x_tx_t')^{-1}(S_{t-1}+\mu
I_p+x_tx_t') \\ &=& \mu^{-1}(I_p+\mu (S_{t-1}+x_tx_t')^{-1}) \\ &=&
 \mu^{-1}I_p+(S_{t-1}+x_tx_t')^{-1} \\ &=& S_{t-1}^{-1} -
\frac{S_{t-1}^{-1}x_tx_t'S_{t-1}^{-1}}{x_t'S_{t-1}^{-1}x_t+1}+\mu^{-1}
I_p \\ &=& R_t-Q_tK_tK_t'+V_{\omega},
\end{eqnarray*}
which is the KF recursion \eqref{Rt:1}, where
$V_{\omega}=\mu^{-1}I_p$.

Clearly, the FLS estimator $\widehat{\beta}_t$ of \eqref{eq:Betat_1}
is the same as the KF estimator $\widehat{\beta}_t$ of
\eqref{beta:est:ss}. From this equivalence, and in particular from
$V_{\omega}=\mu^{-1}I_p$, it follows that
$$
\cov(\beta_{t+1}-\beta_t)=\frac{1}{\mu}I_p
$$

This result further clarifies the role of the smoothing parameter
$\mu$ in \eqref{original_cost}. As $\mu\rightarrow\infty$, the
covariance matrix of $\beta_{t+1}-\beta_t$ is almost zero, which
means that $\beta_{t+1}=\beta_t$, for all $t$, reducing the model to
a usual regression model with constant coefficients. In the other
extreme, when $\mu\approx 0$, the covariance matrix of
$\beta_{t+1}-\beta_t$ has very high diagonal elements (variances)
and therefore the estimated $\beta_t$'s fluctuate erratically.

An important computational consequence of the established
correspondence between the FLS and the KF is apparent. For each time
$t$, FLS requires the inversion of two matrices, namely
$S_{t-1}+x_tx_t'$ and $S_{t-1}+\mu I_p+x_tx_t'$. However, these
inversions are not necessary, as it is clear by the KF that
$\widehat{\beta}_t$ can be computed by performing only matrix
multiplications. This is particulary useful for temporal data mining
data applications when $T$ can be infinite and $p$ very large.

It is interesting to note how the two procedures arrive to the same
solution, although they are based on quite different principles. On
one hand, FLS merely solves an optimization problem, as it minimizes
the cost function $C(\mu)$ of (\ref{original_cost}). On the other
hand, KF performs two steps: first, all linear estimators are
restricted to forms of (\ref{beta:est:ss}), for any parameter vector
$K_t$; in the second step, $K_t$ is optimized so that it minimizes
$P_t$, the covariance matrix of $\beta_t-\widehat{\beta}_t$. This
matrix,  known as the {\it error matrix} of $\beta_t$, gives a
measure of the uncertainty of the estimation of $\beta_t$.

The relationship between FLS and KF has important implications for
both methods. For FLS, it suggests that the regression coefficients
can be learned from the data in a recursive way without the need of
performing matrix inversions; also, the error matrix $P_t$ is
routinely available to us. For KF, we have proved that the estimator
$\widehat{\beta}_t$ minimizes the cost function
$C(\mu)=C(1/V_\omega)$ when only the mean and the variance of the
innovations $\epsilon_t$ and $\omega_t$ are specified, without
assuming these errors to be normally distributed.

\section{An FLS-based algorithmic trading system} \label{sec:system}

\subsection{Data description}

\begin{figure}
\centering
\includegraphics[width=10cm, height=7cm]{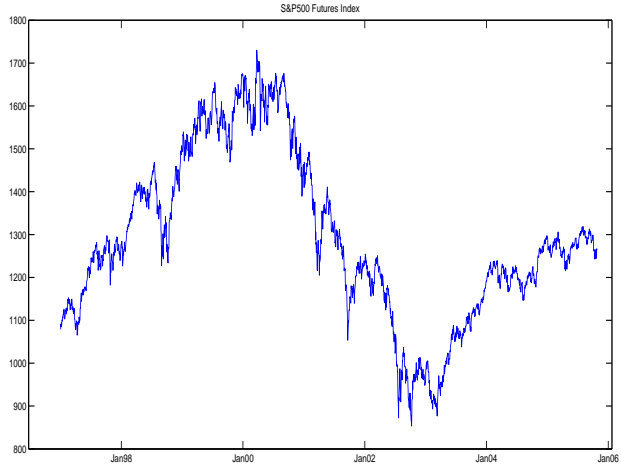}
\caption{{\small S\&P $500$ Futures Index for the available $9$-years period}} \label{fig:index}
\end{figure}

We have developed a statistical arbitrage system that trades S\&P $500$ stock-index futures contracts. The underlying
instrument in this case is the S\&P $500$ Price Index, a world renowned index of $500$ US equities with minimum
capitalization of \$4 billion each; this index is a leading market indicator, and is often used as a gauge of
portfolio performance. The constituents of this index are highly traded by traditional asset management firms and
proprietary desks worldwide. The data stream for the S\&P $500$ Futures Index covers a period of about $9$ years, from
02/01/1997 to 25/10/2005. The contract prices were obtained from Bloomberg, and adjusted\footnote{Futures contracts
expire periodically; since the data for each contract lasts only a few weeks or months, continuous data adjustment is
needed in order to obtain sequences of price data from sequences of contract prices.} to obtain the target data stream
as showed in Figure \ref{fig:index}. Our explanatory data streams are taken to be a subset of all constituents of the
underlying S\&P $500$ Price Index. The constituents list was acquired from the Standard \& Poor's web site as of $1$st
of March 2007, whereas the constituents data streams were downloaded from Yahoo! Financial. The constituents of the
S\&P index are added and deleted frequently on the basis of the characteristics of the index. For our experiments, we
have selected a time-invariant subset of $432$ stocks, namely all the constituents whose historical data is available
over the entire $1997-2005$ period.

The system thus monitors $433$ co-evolving data streams comprising
one target asset and $432$ explanatory streams. All raw prices are
pre-processed in several ways: data adjustments are made for
discontinuities relating to stock splits, bonus issues, and other
financial events; missing observations are filled in using the most
recent data points; finally, prices are transformed into
log-returns. At each time $t>1$, the log-return for asset $i$ is
defined as
$$
r_{it}=\log p_{it}- \log p_{i(t-1)} \qquad i=1,\ldots,432
$$
where $p_{it}$ is the observed price of asset $i$ at time $t$.
Taking returns provides a more convenient representation of the
assets, as it makes different prices directly comparable and center
them around zero. We collect all explanatory assets available at
time $t$ in a column vector $r_t$. Analogously, we denote by $a_t$
the log-return of the S\&P $500$ Futures Index at time $t$.

\subsection{Incremental SVD for dimensionality reduction} \label{sec:svd}

When the dimensionality of the regression model is large, as in our
application, the model might suffer from multicollinearity.
Moreover, in real-world trading applications using high frequency
data, the regression model generating trading signals need to be
updated quickly as new information is acquired. A much smaller set
of explanatory streams would achieve remarkable computational
speed-ups. In order to address all these issues, we implement
on-line feature extraction by reducing the dimensionality in the
space of explanatory streams.

Suppose that  $R_t=\text{E}(r_t r_t')$ is the the unknown population
covariance matrix of the explanatory streams, with data available up
to time $t=1,\ldots,T$. The algorithm proposed by \cite{Weng2003}
provides an efficient procedure to incrementally update the
eigenvectors of the $R_t$ matrix as new data are made available at
time $t+1$. In turn, this procedure allows us to extract the first
few principal components of the explanatory data streams in real
time, and effectively perform incremental dimensionality reduction.

A brief outline of the procedure suggested by  \cite{Weng2003} is
provided in the sequel. First, note that the eigenvector $g_t$ of
$R_t$ satisfies the characteristic equation
\begin{equation}\label{char_equation}
h_t=\lambda_t g_t = R_t g_t
\end{equation}
where $\lambda_t$ is the corresponding eigenvalue. Let us call
$\widehat{h}_t$ the current estimate of $h_t$ using all the data up
to time $t$ $(t=1,\ldots,T)$. We can write the above characteristic
equation in matrix form as
$$
h=\left(\begin{array}{c} h_1 \\ \vdots \\ h_T \end{array}\right) =
\left(\begin{array}{ccc} R_1 & \cdots & 0 \\ \vdots & \ddots &
\vdots \\ 0 & \cdots & R_T \end{array}\right) \left(
\begin{array}{c} g_1 \\ \vdots \\ g_T\end{array}\right) = Rg
$$
and then, noting that
$$
\frac{h_1+\cdots+h_T}{T}=\frac{1}{T}(1,\ldots,1)'h=\frac{1}{T}(R_1,\ldots,R_T)g
= \frac{1}{T}\sum_{i=1}^T R_ig_i
$$
the estimate $\widehat{h}_T$ is obtained by
$\widehat{h}_T=(h_1+\cdots+h_T)/T$ by substituting $R_i$ by
$r_ir_i'$. This leads to
\begin{equation} \label{pca_first}
\widehat{h}_t = \frac{1}{t} \sum_{i=1}^t r_i r_i' g_i
\end{equation}
which is the incremental average of $r_ir_i'g_i$, where $r_ir_i'$
accounts for the contribution to the estimate of $R_i$ at point $i$.

Observing that $g_t=h_t/||h_t||$, an obvious choice is to estimate
$g_t$ as $\widehat{h}_{t-1}/||\widehat{h}_{t-1}||$; in this setting,
$\widehat{h}_0$ is initialized by equating it to $r_1$, the first
direction of data spread. After plugging in this  estimator in
\eqref{pca_first}, we obtain
\begin{equation} \label{eq:ht} h_t = \frac{1}{t} \sum_{i=1}^t
r_i r_i' \frac{\widehat{h}_{i-1}}{||\widehat{h}_{i-1}||}
\end{equation}

In a on-line setting, we need a recursive expression for
$\widehat{h}_t$. Equation \eqref{eq:ht} can be rearranged to obtain
an equivalent expression that only uses $\widehat{h}_{t-1}$ and the
most recent data point $r_t$,
\begin{equation*} \label{pca_second}
\widehat{h}_t = \frac{1}{t} \sum_{i=1}^{t-1} r_ir_i'
\frac{\widehat{h}_{i-1}}{||\widehat{h}_{i-1}||}+\frac{1}{t}r_tr_t'\frac{\widehat{h}_{t-1}}{||\widehat{h}_{t-1}||}=
\frac{t-1}{t} \widehat{h}_{t-1} + \frac{1}{t} r_t r_t'
\frac{\widehat{h}_{t-1}}{|| \widehat{h}_{t-1} ||}
\end{equation*}
The weights $(t-1)/t$ and $1/t$ control the influence of old values
in determining the current estimates. Full details related to the
computation of the subsequent eigenvectors can be found in the
contribution of \cite{Weng2003}.

In our application, we have used data points from 02/01/1997 till
01/11/2000 as a training set to obtain stable estimates of the first
few dominant eigenvectors. Therefore, data points prior to
01/11/2000 will be excluded from the experimental results.

\subsection{Trading rule}

The trade unit for S\&P $500$ Futures Index is set by the Chicago
Mercantile Exchange (CME) to $\$250$ multiplied by the current S\&P
$500$ Price Index, $p_t$. Accordingly, we denote the trade unit
expressed in monetary terms as $C_t=250~p_t$, which also gives the
contract value at time $t$. For instance, if the current stock index
price is $1400$, then an investor is allowed to trade the price of
the contract, i.e. $\$35000$, and its multiples. In our application,
we assume an initial investment of $\$100$ million, denoted by $w$.
The numbers of contracts being traded on a daily basis is given by
the ratio of this initial endowment $w$ to the price of the contract
at time $t$, and is denoted by $\pi_t$.

We call $r_t$ the set of explanatory streams. In the experimental
results of Section \ref{sec:results}, $r_t$ will either be the
$432$-dimensional column vector including the entire set of
constituents (the \emph{without SVD} case), or the reduced
$3$-dimensional vector of three principal components computed
incrementally from the $432$ streams (the \emph{with SVD} case)
using the method of Section \ref{sec:svd}.

Given target and explanatory streams, respectively $a_t$ and $r_t$,
the FLS algorithm updates the current estimate of the artificial
asset at time $t$. With the most updated estimate of the artificial
asset, the current risk (i.e. the regression residual) data point is
derived as
\begin{equation} \label{eq:risk}
s_{t}=a_t - r'_{t}\beta_{t}
\end{equation}

The current position, i.e. the suggested number of contracts to hold
at the end of the current day, is obtained by using
\begin{equation*} \label{eq:contracts}
\vartheta_{t}(s_t)= \phi(\widehat s_{t+1}) \pi_t
\end{equation*}
where $\phi(\widehat s_{t+1})$ is a function of the predicted risk.
In our system, we deploy a simple functional (commonly known to
practitioners as the \emph{plus-minus one} rule), given by
\begin{equation} \label{eq:rule} \phi(\widehat s_{t+1}) =
-\text{sign}(s_t)
\end{equation}

\begin{figure}
\centering
\includegraphics[width=10cm, height=7cm]{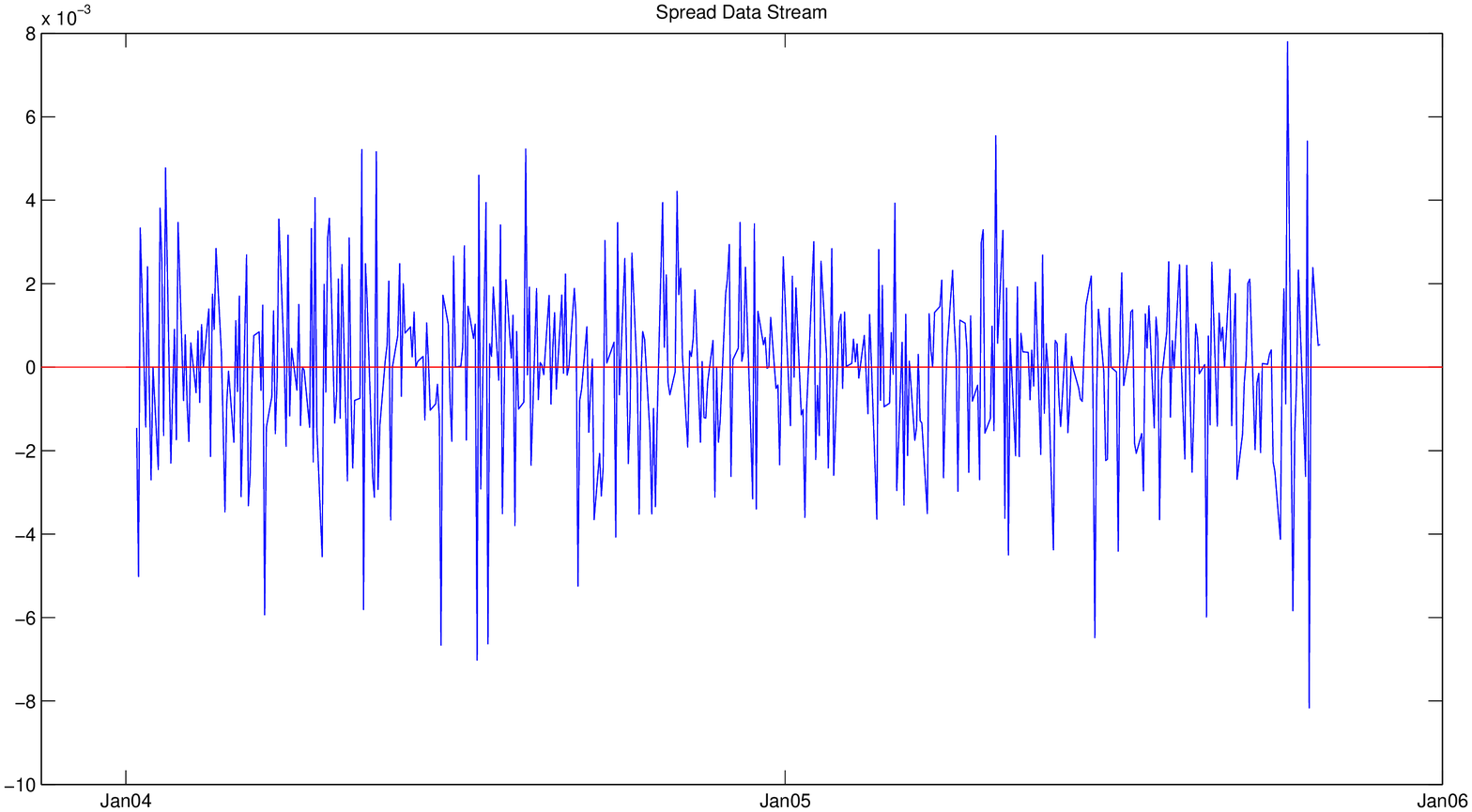}
\caption{{\small Spread stream $s_t$ for a subset of the entire
period. The FLS model is based on the largest principal component
and $\delta=0.2$.}} \label{fig:resid}
\end{figure}

This rule implies that the risk data stream exhibits a
mean-reverting behavior. The spread stream of Figure
\ref{fig:resid}, as well as our experimental results, suggest that
this assumption generally holds true. More formal statistical
procedures could be used instead to test whether mean-reversion is
satisfied at each time $t$. More realistic trading rules would also
be able to detect more general patterns in the spread stream, and
should take into consideration the uncertainty associated with the
presence of such patterns, as well the history of previous trading
decisions.

Having obtained the number of contracts to hold, the daily order
size is given by
$$
\varphi_t= \vartheta_{t}(s_t)- \vartheta_{t-1}(s_t)
$$
rounded to the nearest integer. The trading systems buys or sells
daily in order to maintain the suggested number of contracts. The
monetary return realized by the system at each time $t$ is given by
$$
f_t = 250~(p_t - p_{t-1})~\vartheta_{t-1}(s_t)
$$
%$$
%f_t = r_{0t} \vartheta_{t-1}(s_t) C_{t-1}
%$$

\section{Experimental results} \label{sec:results}

In this section we report on experimental results obtained from the
simple FLS-based trading system. We have tested the system using a
grid of values for the smoothing parameter $\delta$ described in
Section \ref{fls}, to understand the effect of its specification.
Table \ref{table:summaries} shows a number of financial performance
indicators, as well as a measure of goodness of fit, with and
without incremental SVD.

The most important financial indicator is the \emph{Sharpe ratio},
defined as the ratio between the average monetary returns and its
standard deviation. It gives a measure of the mean excess return per
unit of risk; values greater than $0.5$ are considered very
satisfactory, given that our strategy trades one single asset only.
Another financial indicator reported here is the \emph{maximum
drawdown}, the largest movement from peak to bottom of the
cumulative monetary return, reported as percentage. The mean square
error (MSE) has been computed both in sample and out of sample.

% \begin{table}[tbp]
\begin{sidewaystable}
\centering {\tiny
\begin{tabular}{|l||rr|rr|rr|rr|rr|rr|rr|rr|rr|rr|rr|}
\hline
%\multicolumn{21}{|c|}{\textbf{Summary statistics with and without incremental SVD }}\\
\hline $\delta$ & \multicolumn{2}{|c|}{\% gain} &
\multicolumn{2}{|c|}{\% loss} &  \multicolumn{2}{|c|}{MDD} &
\multicolumn{2}{|c|}{\% WT} &\multicolumn{2}{|c|}{\% LT}&
\multicolumn{2}{|c|}{Ann.R.} &\multicolumn{2}{|c|}{Ann.V.}&
\multicolumn{2}{|c|}{Sharpe}
& \multicolumn{2}{|c|}{in-MSE$^*$} & \multicolumn{2}{|c|}{out-MSE$^*$} \\%& \multicolumn{2}{|c|}{MSE-IN/OUT}\\
\hline
$0.01$&$0.786$&$0.773$&$-0.802$&$-0.817$&$31.809$&$28.529$&$47.886$&$48.732$&$43.659$&$42.813$&$ 6.559$&$ 6.728$&$16.393$&$16.393$&$0.400$&$ 0.410$&$0.159$&$0.019$&$2.328$&$2.311$\\
$0.10$&$0.797$&$0.788$&$-0.789$&$-0.799$&$31.569$&$38.770$&$48.194$&$46.887$&$43.351$&$44.658$&$10.610$&$ 3.118$&$16.384$&$16.397$&$0.648$&$ 0.190$&$0.153$&$0.003$&$2.270$&$2.329$\\
$0.20$&$0.803$&$0.792$&$-0.783$&$-0.795$&$28.616$&$34.777$&$48.501$&$46.810$&$43.044$&$44.735$&$13.175$&$ 3.739$&$16.377$&$16.396$&$0.804$&$ 0.228$&$0.149$&$0.001$&$2.243$&$2.333$\\
$0.30$&$0.801$&$0.782$&$-0.785$&$-0.805$&$26.645$&$31.541$&$48.578$&$46.964$&$42.967$&$44.581$&$13.080$&$ 2.115$&$16.377$&$16.398$&$0.799$&$ 0.129$&$0.147$&$0.000$&$2.229$&$2.335$\\
$0.40$&$0.797$&$0.789$&$-0.789$&$-0.798$&$30.201$&$28.432$&$48.117$&$46.887$&$43.428$&$44.658$&$10.287$&$ 3.365$&$16.385$&$16.397$&$0.628$&$ 0.205$&$0.144$&$0.000$&$2.221$&$2.336$\\
$0.50$&$0.788$&$0.789$&$-0.800$&$-0.798$&$29.608$&$29.157$&$48.424$&$46.887$&$43.121$&$44.658$&$ 9.253$&$ 3.356$&$16.388$&$16.397$&$0.565$&$ 0.205$&$0.142$&$0.000$&$2.214$&$2.336$\\
$0.60$&$0.789$&$0.788$&$-0.799$&$-0.800$&$30.457$&$32.752$&$48.655$&$46.656$&$42.890$&$44.889$&$10.381$&$ 2.139$&$16.385$&$16.398$&$0.634$&$ 0.130$&$0.140$&$0.000$&$2.210$&$2.337$\\
$0.70$&$0.787$&$0.781$&$-0.801$&$-0.806$&$30.457$&$36.569$&$48.886$&$46.272$&$42.660$&$45.273$&$10.819$&$-0.950$&$16.384$&$16.398$&$0.660$&$-0.058$&$0.137$&$0.000$&$2.206$&$2.337$\\
$0.80$&$0.789$&$0.782$&$-0.798$&$-0.806$&$33.208$&$34.217$&$48.732$&$46.580$&$42.813$&$44.965$&$10.794$&$ 0.490$&$16.384$&$16.398$&$0.659$&$ 0.030$&$0.134$&$0.000$&$2.202$&$2.338$\\
$0.90$&$0.791$&$0.786$&$-0.796$&$-0.801$&$36.795$&$32.828$&$48.194$&$46.503$&$43.351$&$45.042$&$ 9.074$&$ 1.144$&$16.388$&$16.398$&$0.554$&$ 0.070$&$0.128$&$0.000$&$2.199$&$2.338$\\
$0.99$&$0.800$&$0.787$&$-0.787$&$-0.800$&$32.782$&$33.773$&$47.809$&$46.580$&$43.736$&$44.965$&$ 9.587$&$ 1.689$&$16.387$&$16.398$&$0.585$&$ 0.103$&$0.102$&$0.000$&$2.205$&$2.338$\\

\hline
\end{tabular}
\caption{{\small Experimental results obtained using the statistical
arbitrage system of Section  \ref{sec:system} on $9$-years of S\&P
$500$ Future Index. Each column contains
 a summary statistics obtained \emph{with} (left-hand values) and \emph{without} (right-hand values) incremental SVD.
 The summaries are: daily percentage gain, daily percentage loss, maximum drawdown in percentage, percentage of winning
 trades, percentage of losing trades, annualized percentage return, annualized percentage volatility of returns,
 Sharpe ratio (defined as the ratio of the two previous quantities), in-sample MSE and out-sample MSE.
 *To be multiplied by $10e5$.}}\label{table:summaries}}
\end{sidewaystable}

\begin{figure}
\centering
\includegraphics[width=10cm, height=7cm]{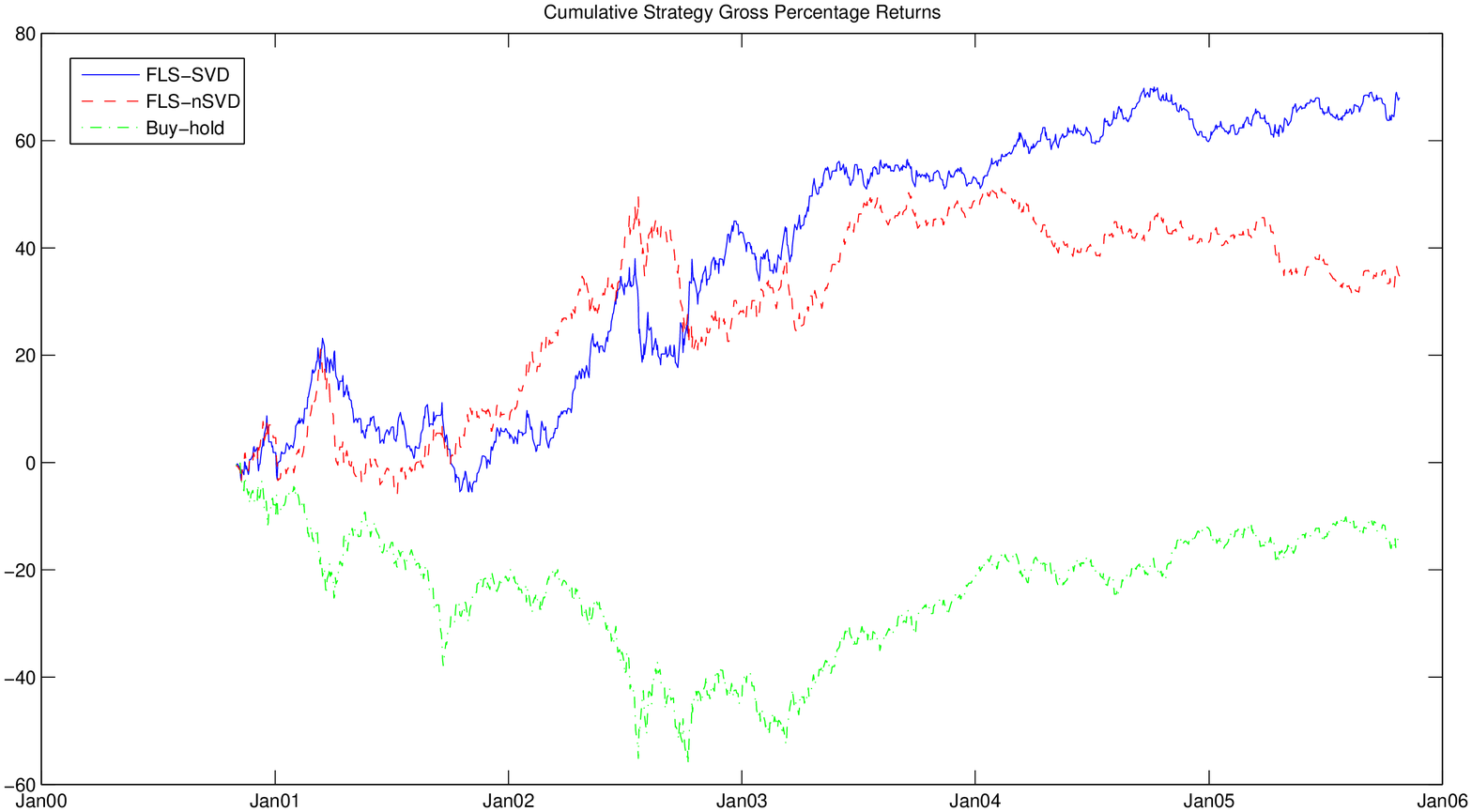}
\caption{{\small Gross profits and losses for three competing
systems: FLS based on SVD (using $\delta=0.2$), FLS based on all
explanatory streams (using $\delta=0.2$) and a buy-and-hold
strategy. }} \label{fig:profits}
\end{figure}

Figure \ref{fig:profits} shows gross percentage returns over the
initial endowment for the constituent set, $f_t/w$, obtained using
three different systems: FLS-based system with incremental SVD
(using only the largest principal component), FLS-based system
without SVD, and a buy-hold strategy. Buy-hold strategies are
typical of asset management firms and pension funds; the investor
buys a number of contracts and holds them throughout the investment
period in question. Clearly, the FLS-based systems outperforms the
index and make a steady gross profit over time. The assumption of
non existence of transactions costs, although simplistic, is not
particularly restrictive, as we expect that this strategy will not
be dominated by cost, given that new transactions are made only
daily. Moreover, we assume that the initial endowment remains
constant throughout the back-testing period, which has an economic
meaning that the investor/agent consumes any capital gain, as soon
as is earned.

\begin{figure}
\centering
\includegraphics[width=10cm, height=7cm]{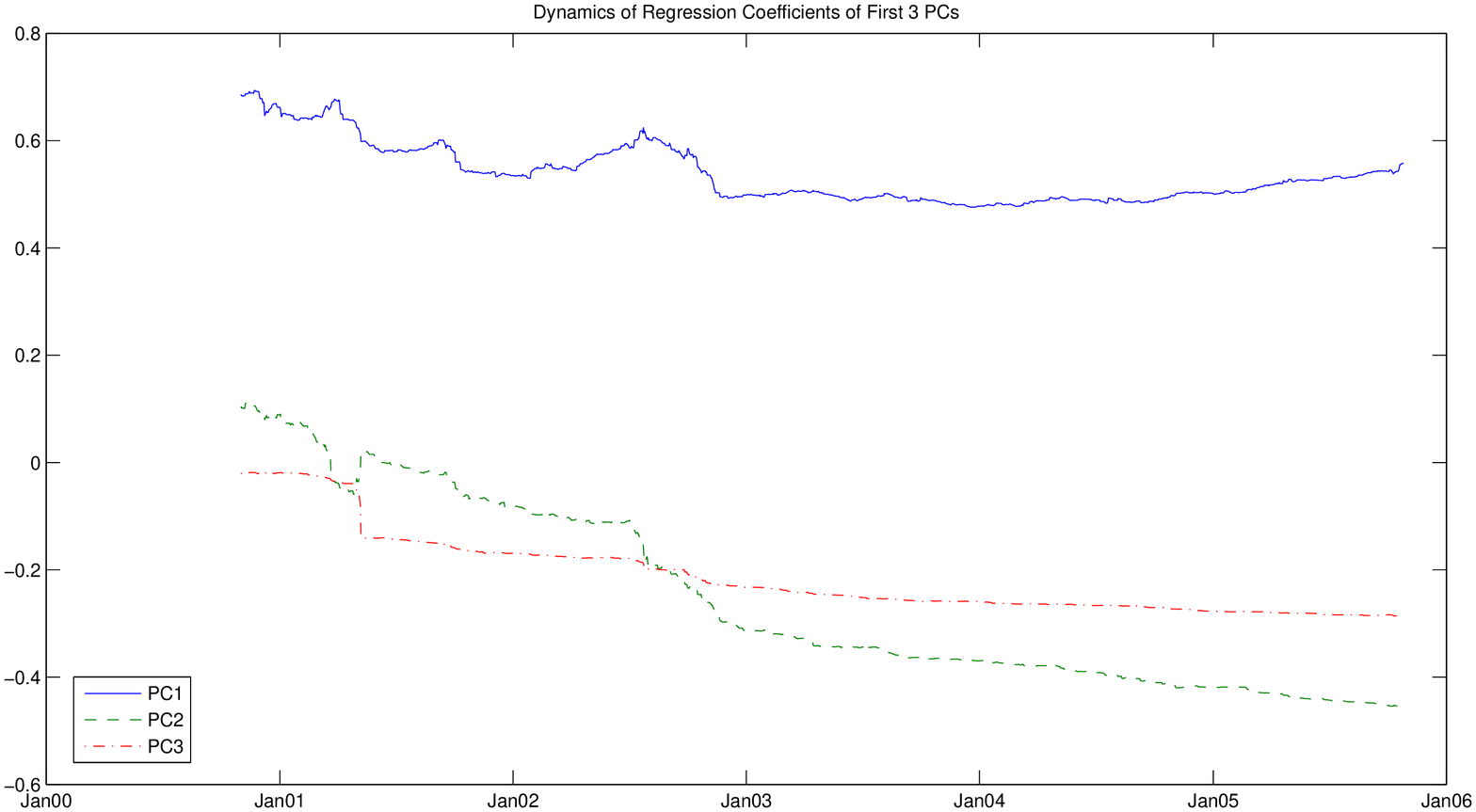}
\caption{{\small Dynamycs of FLS-estimated regression coefficients
associated to the first three principal components, with
$\delta=0.2$. }} \label{fig:betas}
\end{figure}

\begin{figure}
\centering
\includegraphics[width=10cm, height=7cm]{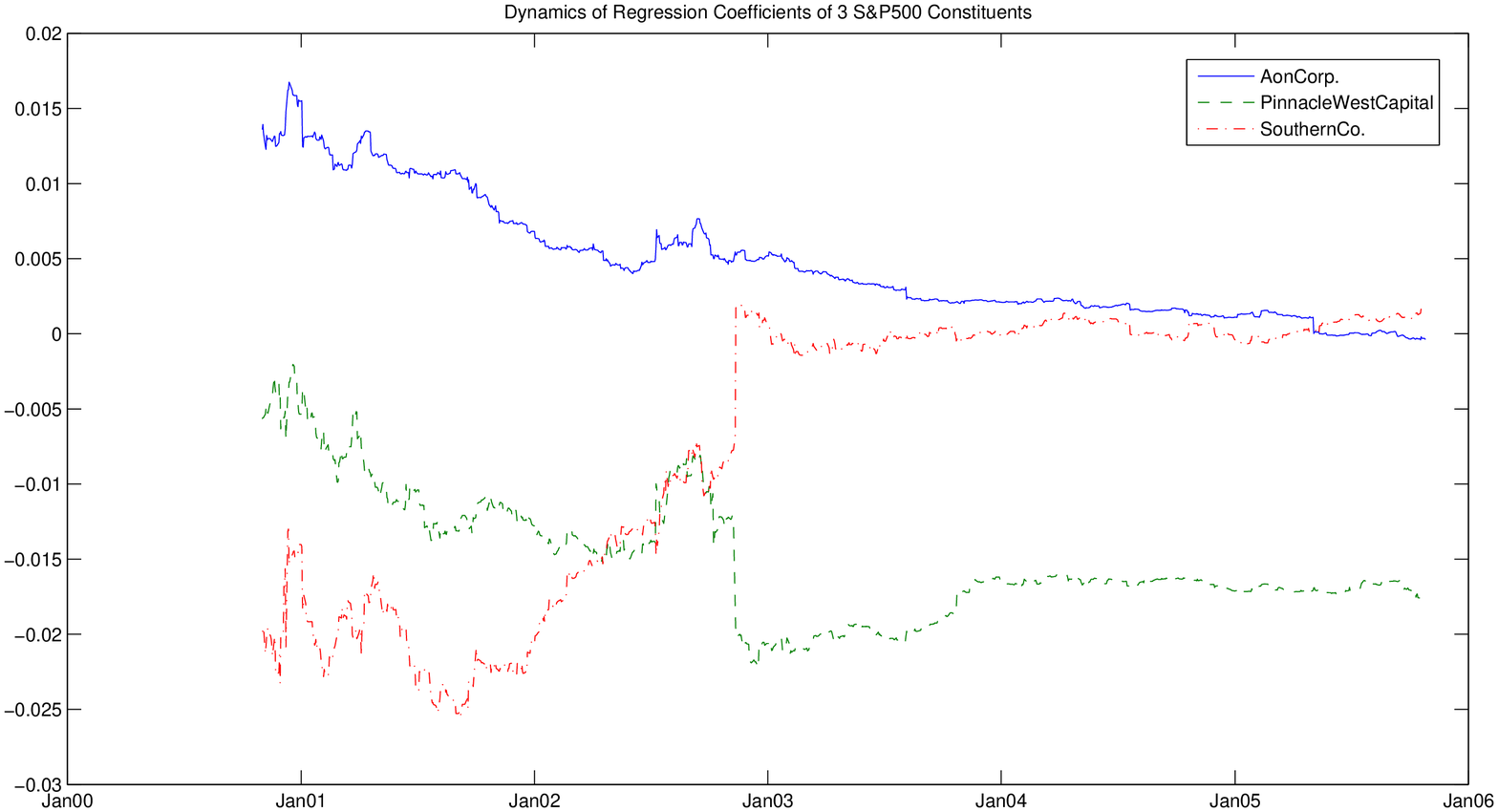}
\caption{{\small Dynamycs of FLS-estimated regression coefficients
associated to three constituents of the index, with $\delta=0.2$. }}
\label{fig:betas2}
\end{figure}

Finally, Figure \ref{fig:betas} shows the estimated time-varying
regression coefficients of the three first principal components, and
Figure \ref{fig:betas2} shows coefficients of three constituent
assets when no SVD has been applied. The coefficients associated to
the first component change very little over the $9$ years period,
whereas the coefficients for the two other components smoothly
decrease over time, with some quite abrupt jumps in the initial
months of $2001$. As we can see from Table \ref{table:summaries}, a
fairly large value of $\delta=0.2$ gives optimal results and
reinforces the merits of time-varying regression in this context.

\begin{figure}
\centering
\includegraphics[width=10cm, height=7cm]{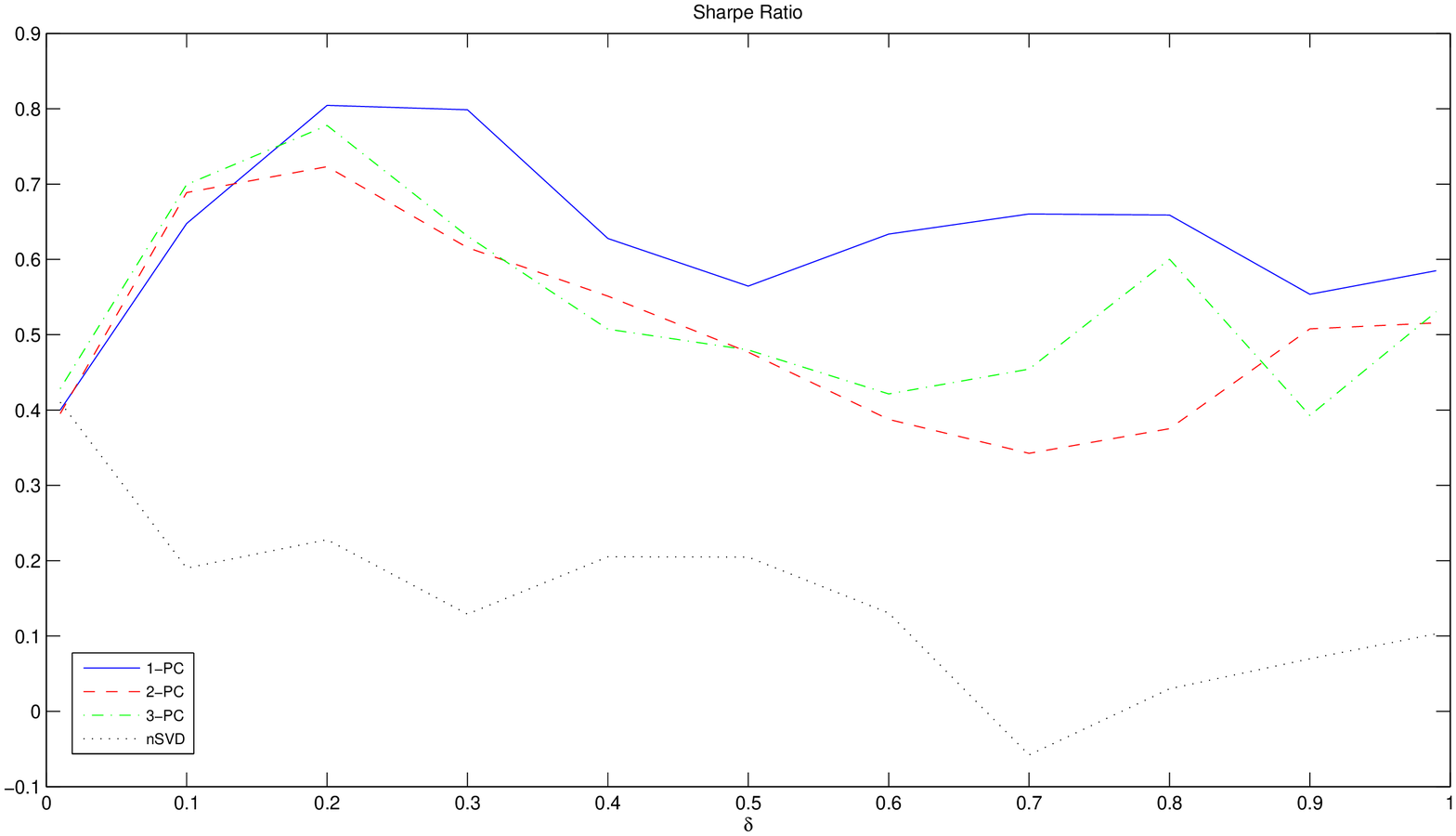}
\caption{{\small Sharpe ratio as function of $\delta$ }} \label{fig:sharpe}
\end{figure}

\section{Conclusions} \label{conclusions}

We have argued that the FLS method for regression with time-varying
coefficients lends itself to a useful temporal data mining tool. We
have derived a clear connection between FLS and Kalman filter
equations, and have demonstrated how this link enhances
interpretation of the smoothing parameter featuring in cost function
that FLS minimizes, and naturally leads to a more efficient
algorithm. Finally, we have shown how FLS can be employed as a
building-block of an algorithmic trading system.

There are several aspects of the simple system presented in Section
\ref{sec:system} that can be further improved upon, and the
remainder of this discussion points to a few general directions and
related work that we intend to explore in the future.

The problem of feature selection is an important one. In Section
\ref{sec:system} the system relies on a set of $432$ constituents of
the S\&P $500$ Price Index under the assumption that they explain
well the daily movements in the target asset. These explanatory data
streams could be selected automatically, perhaps even dynamically,
from a very large basket of streams, on the basis of they
\emph{similarity} to the target asset. This line of investigation
relates to the \emph{correlation detection} problem for data
streams, a well-studied and recurrent issue in temporal data mining.
For instance, \cite{Guha2003} propose an algorithm that aims at
detecting linear correlation between multiple streams. At the core
of their approach is a technique for approximating the SVD of a
large matrix by using a (random) matrix of smaller size, at a given
accuracy level; the SVD is then periodically and randomly
re-computed over time, as more data points arrive. The SPIRIT system
for streaming pattern detection of \cite{Papadimitriou2005} and
\cite{Sun2006} incrementally finds correlations and hidden variables
summarising the key trends in the entire stream collection.

Of course, deciding on what similarity measure to adopt in order to
measure how \emph{close} explanatory and target assets are is not an
easy task, and is indeed a much debated issue (see, for instance,
\cite{Gavrilov2000}). For instance, \cite{Shasha2004} adopt a
sliding window model and the Euclidean distance as a measure of
similarity among streams. Their \emph{StatStream} system can be used
to detect pairs of financial time series with high correlation,
among many available data streams. \cite{Cole2005} combine several
techniques (random projections, grid structures, and others) in
order to compute Pearson correlation coefficients between data
streams. Other measures, such as dynamic time warping, have also
been suggested \citep{Capitani2005}.

Real-time feature selection can be complemented by feature
extraction. In our system, for instance, we incrementally reduce the
original space of $432$ explanatory streams to a handful of
dimensions using an on-line version of SVD. Other dynamic
dimensionality reduction models, such as incremental independent
component analysis \citep{Basalyga2004} or non-linear manifold
learning \citep{Law2004}, as well as on-line clustering methods,
would offer potentially useful alternatives.

Our simulation results have shown gross monetary results, and we
have assumed that transaction costs are negligible. Better trading
rules that explicitly model the mean-reverting behavior (or other
patterns) of the spread data stream and account for transaction
costs, as in \cite{Carcano2005}, can be considered. The trading rule
can also be modified so that trades are placed only when the spread
is, in absolute value, greater than a certain threshold determined
in order to maximize profits, as in \cite{Vidyamurthy2004}. In a
realistic scenario, rather than trading one asset only, the investor
would build a portfolio of models; the resulting system may be
optimized using measures that capture both the forecasting and
financial capabilities of the system, as in \cite{Towers2001}.

Finally, we point out that the FLS method can potentially be used in
other settings and applications, such as predicting co-evolving data
streams with missing or delayed observations, as in \cite{Yi2000},
and for outlier and fraud detection, as in \cite{Adams2006}.

\section*{Acknowledgements}

We would like to thank David Hand for helpful comments on an earlier
draft of the paper.

\bibliographystyle{plainnat}
\bibliography{bibliography}

\end{document}